\documentclass[apj,iop]{emulateapj}
\slugcomment{The Astrophysical Journal, 771:101 (10pp), 2013 July 10} 
\usepackage[table]{xcolor}
\usepackage[normalem]{ulem}
\usepackage{rotating}
\usepackage{tabularx,booktabs}
\shortauthors{Pasham \& Strohmayer}
\begin{document}
\title{On the nature of the \lowercase{m}H\lowercase{z} X-ray QPOs from Ultraluminous X-ray source M82 X-1: Search for Timing-Spectral correlations}
\author{Dheeraj R. Pasham\altaffilmark{1,2}, Tod E. Storhmayer\altaffilmark{2}}
\affil{$^{1}$Astronomy Department, University of Maryland, College Park, MD 20742; email: dheeraj@astro.umd.edu; richard@astro.umd.edu \\ 
$^{2}$Astrophysics Science Division, NASA's GSFC, Greenbelt, MD 20771; email: tod.strohmayer@nasa.gov \\
{\it Received 2013 February 3; accepted 2013 May 17; published 2013 June 21}
}

\begin{abstract}
Using all the archival {\it XMM-Newton} X-ray (3-10 keV) observations of the ultraluminous X-ray source (ULX) M82 X-1 we searched for a correlation between its variable mHz quasi-periodic oscillation (QPO) frequency and its hardness ratio (5-10 keV/3-5 keV), an indicator of the energy spectral power-law index. When stellar-mass black holes (StMBHs) exhibit Type-C low-frequency QPOs ($\sim$0.2-15 Hz) the centroid frequency of the QPO is known to correlate with the energy spectral index. The detection of such a correlation would strengthen the identification of M82 X-1's mHz QPOs as Type-C and enable a more reliable mass estimate by scaling its QPO frequencies to those of Type-C QPOs in StMBHs of known mass. We resolved the count rates and the hardness ratios of M82 X-1 and a nearby bright ULX (source 5/X42.3+59) through surface brightness modeling. We detected QPOs in the frequency range of 36-210 mHz during which M82 X-1's hardness ratio varied from 0.42-0.47. Our primary results are: (1) we do not detect any correlation between the mHz QPO frequency and the hardness ratio (a substitute for the energy spectral power-law index) and (2) similar to some accreting X-ray binaries, we find that M82 X-1's mHz QPO frequency increases with its X-ray count rate (Pearson's correlation coefficient = +0.97). The apparent lack of a correlation between the QPO centroid frequency and the hardness ratio poses a challenge to the earlier claims that the mHz QPOs of M82 X-1 are the analogs of the Type-C low-frequency QPOs of StMBHs. On the other hand, it is possible that the observed relation between the hardness ratio and the QPO frequency represents the saturated portion of the correlation seen in Type-C QPOs of StMBHs -- in which case M82 X-1's mHz QPOs can still be analogous to Type-C QPOs.

\end{abstract}


\vfill\eject

\newpage

\section{Introduction}
The bright, point-like, non-nuclear X-ray sources in nearby galaxies with X-ray (0.3-10.0 keV) luminosities in the range of a few$\times$10$^{39-41}$ ergs s$^{-1}$ are known as the Ultra-Luminous X-ray sources (ULXs). Their variability on short timescales (some ULXs vary on timescales of the order of a few minutes) combined with high X-ray luminosities suggests that these sources are powered by accretion of matter onto black holes (this excludes the X-ray bright supernovae: see, for example, Immler \& Lewin 2003). But the masses of these black holes is still controversial. The current arguments suggest that ULXs are either powered by stellar-mass black holes (StMBH: mass range of 3-50 M$_{\odot}$) accreting matter via a super-Eddington mechanism (e.g., K\"ording et al. 2002; King et al. 2001; Begelman 2002; Gladstone et al. 2009), or that they comprise Intermediate-Mass Black Holes (IMBHs: mass range of a few$\times$(100-100) M$_{\odot}$) accreting at a sub-Eddington rate (Colbert \& Mushotzky 1999). There is, however, no clear consensus on either scenario.

A subsample of ULXs show X-ray quasi-periodic oscillations (QPOs). These include NGC 5408 X-1 (centroid frequencies of $\approx$ 10-40 mHz: Strohmayer et al. 2007; Strohmayer \& Mushtozky 2009; Dheeraj \& Strohmayer 2012),  NGC 6946 X-1 (centroid frequency of 8.5 mHz: Rao et al. 2010), M82 X-1 (centroid frequencies of $\approx$ 50-170 mHz: Strohmayer \& Mushtozky 2003; Dewangan et al. 2006; Mucciarelli et al. 2006) and X42.3+59 (centroid frequencies of 3-4 mHz: Feng et al. 2010). In particular, the qualitative nature of the power density spectra (PDS) of NGC 5408 X-1, NGC 6946 X-1 and M82 X-1 is similar and can be described by a flat-topped, band-limited noise breaking to a power-law with QPOs evident on the power-law portion of the PDS, close to the break. This behavior is strikingly similar to the PDS of StMBHs, when they exhibit the so-called ``type-C'' Low-Frequency QPOs (LFQPOs: frequency range of $\approx$ 0.2-15 Hz). However, the crucial difference is that the characteristic frequencies within the PDS of the ULXs, viz., the break frequency and the centroid frequency of the QPOs are scaled down by a factor of approximately 10-100 compared to the StMBHs with type-C LFQPOs. It has thus been argued that the mHz QPOs (10-200 mHz) of ULXs are the analogs of the type-C LFQPOs of StMBHs and that the observed difference in the characteristic frequencies (a few$\times$(0.01-0.1) Hz compared with a few Hz) is due to the presence of massive black holes ($>$ mass of the StMBHs) within the ULX systems. 

Furthermore, it has been established recently (e.g., McHardy et al. 2006; K{\"o}rding et al. 2007) that the break frequency of the PDS of StMBHs and super-massive black holes scales inversely with the mass of the black hole (after accounting for the differences in the luminosities, i.e., accretion rate of the sources). In addition, it is known that the centroid frequency of the LFQPOs of StMBHs scales directly with the break frequency of the PDS (Wijnands \& van der Klis 1999; Klein-Wolt \& van der Klis 2008). Therefore, it is reasonable to assume that the centroid frequency of the type-C LFQPOs and its analogs, if any, in ULXs \& AGN should scale with the mass of the host black hole. However, the LFQPOs of StMBHs are variable and occur in a wide range of frequencies ($\approx$ 0.2-15 Hz). But, combining spectral information has proven to be useful. One of the distinctive features of the type-C LFQPOs of StMBHs is that their variable centroid frequency is strongly correlated with the index of the power-law component of the energy spectrum. The trend can be described as an increase in the power-law index with increase in the centroid frequency of the QPO with an evidence for either a turn-over or saturation, i.e., decrease or constancy of the power-law index with increase in the QPO centroid frequency, beyond a certain high QPO frequency (see Figure 10 of Vignarca et al. 2003). Therefore, at a given value of the energy spectral power-law index, the QPO frequency scales directly with the mass of the black hole. Hence, under the assumption that a mHz QPO from a certain ULX is an analog of the type-C LFQPOs of StMBHs, its black hole mass can be estimated by measuring the QPO frequency from the PDS and the power-law index from its energy spectrum.

This method of constraining the masses of ULXs has been employed by
several authors. Dewangan et al. (2006) constructed the PDS of M82 X-1
using the longest available {\it XMM-Newton} observation and detected
a QPO with a centroid frequency of $\approx$ 114 mHz. During this
observation they found that the index of the power-law component of
the energy spectrum was $\approx$ 2. Assuming that these two
quantities correlate in a similar manner to that observed in the StMBH
binaries XTE J1550-564 and GRS 1915+105, they obtained a mass range of
25-500 M$_{\odot}$ from scaling the respective QPO centroid
frequencies. In the same way, Rao et al. (2010) estimated a mass range
of (1-4)$\times$1000 M$_{\odot}$ for the black hole in NGC 6946
X-1. Finally, Strohmayer \& Mushotzky (2009) found that both the PDS
and the X-ray energy spectra of NGC 5408 X-1 are qualitatively similar to
those of StMBHs when they are in the so-called steep power-law (SPL)
accretion state. Using the QPO centroid frequency -- energy spectral
power-law index relations of five reference stellar-mass black holes
in the SPL state, they used QPO frequency scaling to estimate a mass
of $\sim$ a few$\times$1000 M$_{\odot}$ for NGC 5048 X-1.

It is crucial to realize that all the current black hole mass estimates of ULXs, that rely on scaling QPO frequencies at a given power-law index, assume that the mHz QPOs seen in ULXs are the analogs of the type-C LFQPOs of StMBHs. In this article, we test this hypothesis in the case of ULX M82 X-1, by investigating if its QPOs show the same characteristic behavior of type-C LFQPOs of StMBHs, i.e., {\it whether M82 X-1's QPO frequency is correlated with the power-law index of its energy spectrum}. Similar attempts have been made earlier by Fiorito \& Titarchuk (2004) for the case of M82 X-1 and more recently by Dheeraj \& Strohmayer (2012) for the case of NGC 5408 X-1. The work by Fiorito \& Titarchuk (2004) considered only one {\it XMM-Newton} observation and three {\it RXTE/PCA} observations and was severely limited by the observed variability of M82 X-1's QPO frequencies, i.e, 50-100 mHz. In addition, they did not consider the contamination by a nearby bright X-ray source (source 5/X42.3+59) in their spectral modeling. Here we include analysis using all of the archival {\it XMM-Newton} observations that show QPOs in the frequency range of 36 mHz (the lowest ever reported from M82 X-1) to 210 mHz (the highest QPO frequency reported from M82 X-1). 

This article is arranged as follows. In Section 2, we describe all the {\it XMM-Newton} observations used in the present study and carry out surface brightness modeling of their MOS1 images. In Section 3, we show results from our timing and energy-dependent surface brightness modeling analysis. We also show the two primary results of this article: (1) evidence for a correlation between the average count rate and the centroid frequency of the QPO and (2) no apparent correlation between the centroid frequency of the QPO and the hardness ratio which is an indicator of the power-law index of the energy spectrum. In Section 4, we compare these results with StMBHs with type-C QPOs. We discuss the implications of the observed correlations on the mass of the black hole within M82 X-1. 

\begin{table}
    \caption{{ Resolved average count rates (3-10 keV) of M82 X-1 and source 5 derived from the surface brightness modeling of {\it XMM-Newton}'s MOS1 images.}}\label{Table1}
{\small
\begin{center}
   \begin{tabular}[t]{lccc} 
\hline\hline \\
ObsID\tablenotemark{a} & Source 5  & M82 X-1 & $\chi^{2}$/dof\tablenotemark{c} \\
			& (counts s$^{-1}$)\tablenotemark{b} & (counts s$^{-1}$)\tablenotemark{b}  & \\
\\
    \hline \\
  0112290201 & $0.071 \pm 0.003$ & $0.041 \pm 0.003$  & 627/437  \\

  0206080101 & $0.011 \pm 0.002$ & $0.046 \pm 0.002$  & 509/437 \\

 0657800101 & $0.034 \pm 0.003$ & $0.037 \pm 0.003$ & 417/437 \\

 0657801901 & $0.015 \pm 0.002$ & $0.035 \pm 0.002$ & 400/437 \\

 0657802101 & $0.025 \pm 0.003$ & $0.040 \pm 0.003$ & 435/437 \\

 0657802301& $0.047 \pm 0.004$ & $0.053 \pm 0.004$ & 571/437 \\
\\
    \hline
    \end{tabular}
\end{center}
}
\tablenotemark{a}{The {\it XMM-Newton} assigned observation ID.}\\
\tablenotemark{b}{The count rates are calculated using the formula described in the text (see Section 2).}\\
\tablenotemark{c}{The $\chi^2$/dof was obtained by fitting two point spread functions to MOS1 images of size 21''$\times$21'' binned to 1''$\times$1'' and centered on M82 X-1.}\\
\end{table}


\section{ {\it XMM-Newton} observations and surface brightness modeling }
Prior to the present work, QPOs have been reported from M82 X-1 using the {\it RXTE/PCA}  (Kaaret et al. 2006; Mucciarelli et al. 2006) and the {\it XMM-Newton}/EPIC data (Strohmayer \& Mushotzky 2003; Mucciarelli et al. 2006; Dewangan et al. 2006). {\it RXTE}'s PCA is a non-imaging detector whose field of view includes various point sources nearby M82 X-1. Its data does not allow one to disentangle the contribution from the nearby bright sources. However, data acquired with {\it XMM-Newton} allows for surface brightness modeling that can help us understand M82 X-1's relative brightness with respect to the  nearby sources. Also, {\it XMM-Newton} observations have longer exposures which allow firm detection of the QPOs. Due to these reasons, we decided to use only {\it XMM-Newton} data. To date, {\it XMM-Newton} has observed M82 on twelve occasions. Three of these observations were severely effected by flaring. We analyzed the remaining nine observations to search for the presence of QPOs. We detected QPOs in six of them. Since the present work relies on searching for a correlation between the QPO frequency and the energy spectral power-law index, we only considered the observations with QPOs. The {\it XMM-Newton} assigned IDs of the six observations used in this article are 0112290201, 0206080101, 0657800101, 0657801901, 0657802101 and 0657802301. The total observing times are 30 ks, 104 ks, 26 ks, 28 ks, 22 ks and 23 ks, respectively. 

At {\it XMM-Newton}'s spatial resolution the flux from M82 X-1 is contaminated by the diffuse X-ray emission from the host galaxy (e.g., Strickland \& Heckman 2007) and the nearby point sources (Matsumoto et al. 2001). Careful X-ray spectral modeling by various authors including Mucciarelli et al. (2006) and Caballero-Garc{\'{\i}}a (2011) has shown that the diffuse component is dominant at energies below 3 keV. Therefore, to eliminate its contribution, we only included events in the energy range of 3.0-10.0 keV. Similar exclusions have been employed by Strohmayer \& Mushotzky (2003), Fiorito \& Titarchuk (2004) and Dewangan et al. (2006). The observations taken by the high-resolution camera on board {\it Chandra} have revealed that there are a total of nine point sources within the 10''$\times$10'' region around M82 X-1 (Matsumoto et al. 2001). In principle, the flux contribution from all these point sources can bias the modeling of M82 X-1. Chiang \& Kong (2011) have analyzed all of the archival {\it Chandra} observations of M82 to study the long-term (1999-2007) variability of the X-ray point sources within M82. They find that while the X-ray sources nearby M82 X-1 are variable, the maximum observed X-ray (0.3-8.0 keV) luminosity of these sources is $\la$ ${1/5}^{th}$ the average luminosity of M82 X-1 (see Table 2 of Chiang \& Kong 2011). However, source 5 (as defined in Matsumoto et al. 2001) is an exception. It can reach X-ray luminosities comparable to M82 X-1 (Feng \& Kaaret 2007). Therefore, to estimate the amount of contamination by source 5 in each of the observations, we carried out surface brightness modeling of the images assuming they are dominated by two point sources.

We used only the MOS1 data for the purposes of surface brightness modeling. This is due to the fact that the MOS data offers the finest pixel size of 1.1'' compared to the 4.1'' of the EPIC-pn. Furthermore, the image resolution of EPIC-pn is close to the separation ($\approx$ 5'') between source 5 and M82 X-1 (Feng \& Kaaret 2007). We avoid MOS2 data because its point spread function (PSF) is non-axisymmetric at the core. The on-axis PSF of MOS1 can be adequately described by an axisymmetric 2D king model ({\it XMM-Newton} current calibration file release notes 167). Similar to the analysis of Feng \& Kaaret (2007) (who also carried out surface brightness modeling of {\it XMM-Newton}'s MOS1 data of M82 using a king model), we used the {\it calview} tool with an EXTENDED accuracy level to extract an on-axis PSF at an energy of 3.0 keV. We then fit a king model\footnote{\[PSF_{king} = \frac{N}{\left[1+ \left(\frac{r}{r_0} \right)^2\right]^\alpha} \] where r$_{0}$, $\alpha$ and N are the core radius, index and the normalization, respectively.} to this PSF. The best-fit values of the core radius and the index are 4.0'' and 1.39, respectively. We note that these values are consistent with the best-fit parameters given in the latest calibration file XRT1\_XPSF\_0014.CCF and also with the values reported in the MOS calibration documentation ({\it XMM-Newton} current calibration file release notes 167).


\begin{figure*}

\begin{center}
\vspace{-0.75cm}
\includegraphics[width=5.5in, height=6.875in, angle=0]{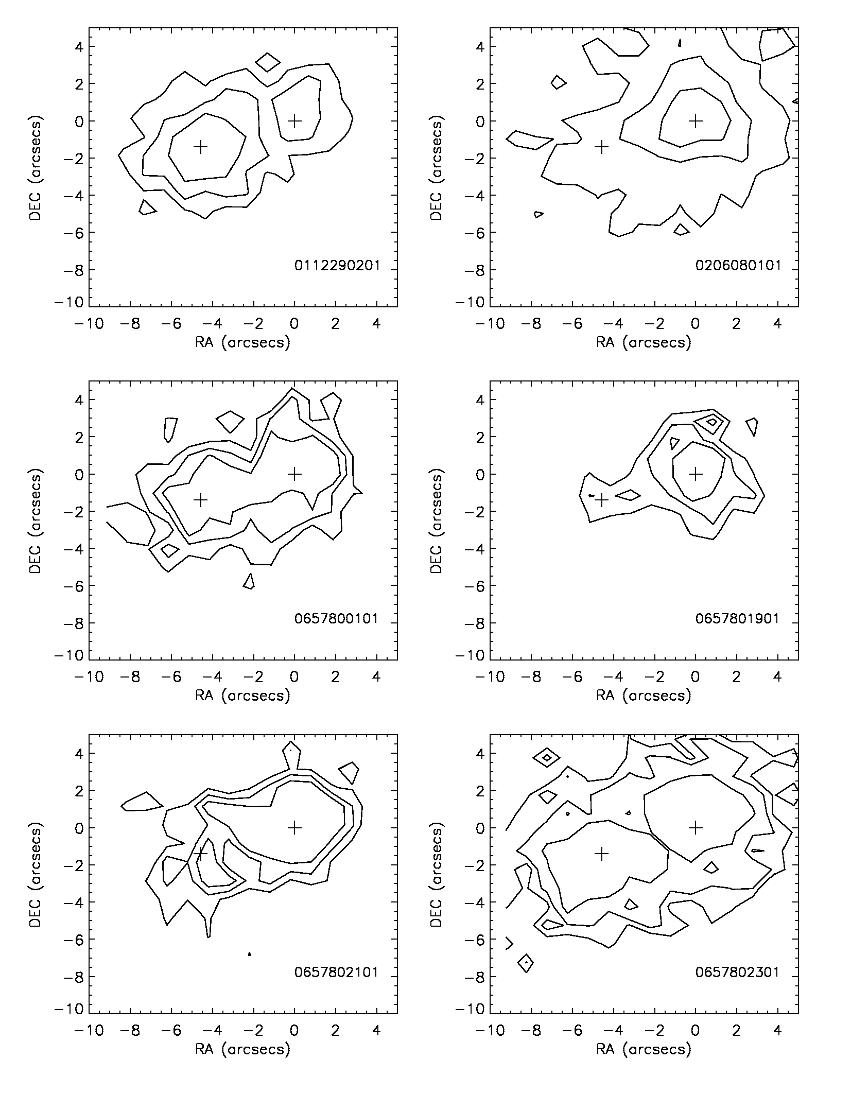}
\end{center}
\vspace{-.35cm}
\caption{Surface brightness contour maps of the MOS1 images (3-10 keV) of M82 during six different epochs. The {\it XMM-Newton} assigned observation IDs are indicated at the bottom right of each panel. M82 X-1 is at the origin in all the plots and the best-fit positions of source 5 and M82 X-1 are represented by plus signs. Contour levels are different for different observations. Top left panel: The contour levels are 1.0, 1.75, 2.5 (10$^{-3}$ counts s$^{-1}$ arcsec$^{-2}$). Top right panel: The contour levels are 0.5, 1.0, 1.5 (10$^{-3}$ counts s$^{-1}$ arcsec$^{-2}$). Middle left panel:  The contour levels are 0.75, 1.0, 1.25 (10$^{-3}$ counts s$^{-1}$ arcsec$^{-2}$). Middle right panel:  The contour levels are 0.75, 1.0, 1.25 (10$^{-3}$ counts s$^{-1}$ arcsec$^{-2}$). Bottom left panel:  The contour levels are 0.75, 1.0, 1.25 (10$^{-3}$ counts s$^{-1}$ arcsec$^{-2}$). Bottom right panel:  The contour levels are 0.75, 1.0, 1.5 (10$^{-3}$ counts s$^{-1}$ arcsec$^{-2}$). }
\label{fig:figure1}
\end{figure*}


From each of the six {\it XMM-Newton} observations, we extracted an exposure-corrected (using XMMSAS task {\it eexpmap}) MOS1 image of size 21''$\times$21'' binned to 1''$\times$1'' (square pixels) and roughly centered on M82 X-1. The standard filters of {\it FLAG==0} and {\it PATTERN$<$=12} were applied. As mentioned earlier, all the images were extracted in the energy range of 3.0-10.0 keV to reduce the influence of the diffuse X-ray emission from the host galaxy. Each of these MOS1 images were then modeled with two PSFs to represent source 5 and M82 X-1. The core radius and the spectral index of the two PSFs were fixed at the best-fit values, i.e., 4.0'' and 1.39, respectively. The centroids ($x$, $y$) and the normalizations of the two PSFs were allowed to vary. However, the distance between the two sources was fixed to the values found using the co-ordinates reported by Feng \& Kaaret (2007). We ignore the background as it was negligible in all of the six observations. For bins with less than 5 counts, we assign error bars as derived by Gehrels (1986), i.e., 1.0 + $\sqrt{counts + 0.75}$; And for bins with greater than 5 counts we assign Poisson errors of $\sqrt{counts}$. The model with two PSFs yielded acceptable values of $\chi^{2}$ in all the six cases. The best-fit $\chi^{2}$ value for each case is reported in the last column of Table 1. It should be noted that the effective exposure of all but observation ID 0206080101 are comparable. The observation length of 0206080101 is $\approx$ 100 ks while that of the rest of the observations is $\approx$ 25 ks. This dataset was also analyzed by Feng \& Kaaret (2007) and they find that the long exposure causes the other dim sources nearby to be significant for surface brightness modeling. Therefore, to be consistent across all the observations we choose data from one of the good time intervals of MOS1 with an effective exposure of 30 ks. This is comparable to the exposure times of the other five observations. The X-ray (3-10 keV) surface brightness contour maps of all of the six observations are shown in Figure 1. 


\begin{figure*}

\begin{center}
\includegraphics[width=6.5in, height=5.75in, angle=0]{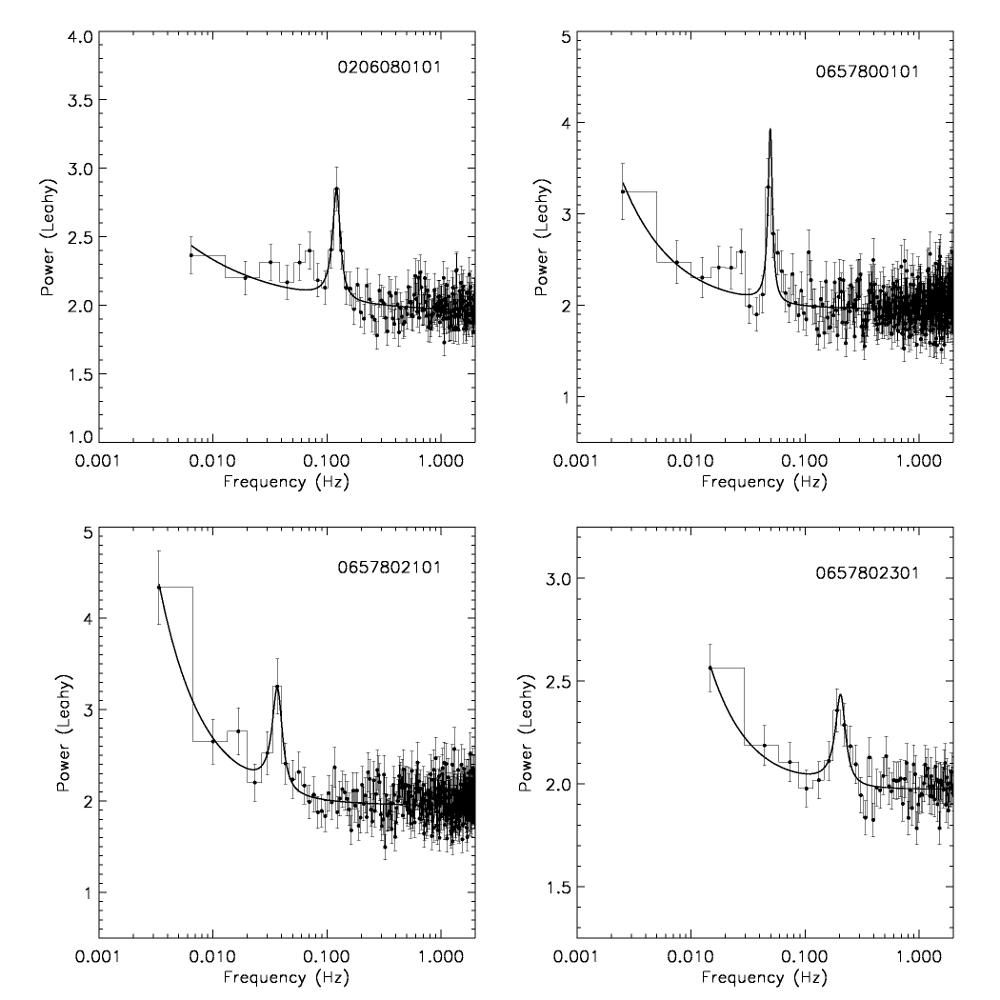}
\end{center}
{\textbf{Figure 2:} The EPIC-pn 3-10 keV power density spectra ({\it histogram}) and the best-fit model ({\it solid}) of four of the five {\it XMM-Newton} observations.  The error bars are also shown. The {\it XMM-Newton} assigned observation IDs are shown on the top right of each panel. }
\label{fig:figure2}
\end{figure*}



\begin{figure*}

\begin{center}
\includegraphics[width=6.25in, height=3.in, angle=0]{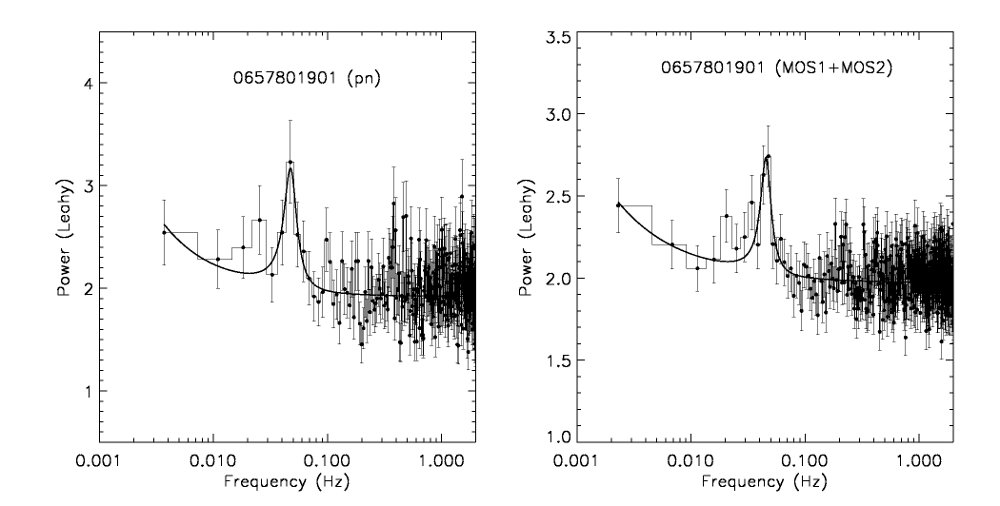}
\end{center}
{\textbf{Figure 3:} Left Panel: The EPIC-pn 3-10 keV power density spectrum ({\it histogram}) and the best-fit model ({\it solid}) of the observation ID 0657800101. Right Panel: The combined EPIC-MOS1 and EPIC-MOS2 3-10 keV power density spectrum ({\it histogram}) and the best-fit model ({\it solid}) of the same observation. In the EPIC-pn data the QPO is significant at only 3$\sigma$ level. However, in the combined MOS power density spectrum the QPO is significant at 5$\sigma$ level. }
\label{fig:figure3}
\end{figure*}



\begin{table*}
  \begin{flushleft}
\centering
  \caption{Summary of the 3-10 keV power spectral modeling.}\label{Table2} 
{\small
    \begin{tabular}[t]{lccccccc}
	\\
    \hline\hline \\
   ObsID 					& 0206080101 & 0657800101 & 0657801901\tablenotemark{c} & 0657801901\tablenotemark{c}     & 0657802101     & 0657802301 \\
						&	     &            &  \hspace{0.15cm}(pn)   & \hspace{0.15cm}(MOS)  &		      &            \\
\\
    \hline \\
Exposure$^{a}$(ks)         	& 60.0       &    22.0    &       8.8 		    & 	24.2 		     & 17.4 	      & 17.0 	    \\
\\
A$^{\ast}$			& $1.94 \pm 0.04$ & $1.96 \pm 0.01$ & $1.92 \pm 0.02$ & $1.97 \pm 0.02$ & $1.95 \pm 0.01$ & $1.97 \pm 0.01$ \\ 
\\
B$^{\ast}$			& $0.03 \pm 0.03$ & $0.01 \pm 0.01$ & $0.01 \pm 0.01$ & $0.01 \pm 0.01$ & $0.01 \pm 0.01$ & $0.01 \pm 0.01$  \\   
\\
$\Gamma$$^{\ast}$		& $0.55 \pm 0.23$ & $1.09 \pm 0.29$ & $0.92 \pm 0.64$ & $0.76 \pm 0.30$ & $1.12 \pm 0.22$ & $1.28 \pm 0.52$   \\ 
\\
N$_{QPO}$$^{\dagger}$		& $0.81 \pm 0.16$ & $1.25 \pm 0.29$ & $1.19 \pm 0.39$ & $0.71 \pm 0.18$ & $1.13 \pm 0.31$ & $0.44 \pm 0.18$  \\
\\
$\nu_{0}$$^{\dagger}$(mHz) 	& $121.4 \pm 2.9$ & $49.3 \pm 1.5$  & $47.4 \pm 2.5$  & $45.4 \pm 1.3$  & $36.7 \pm 2.1$  & $204.8 \pm 6.3$ \\  
\\
$\Delta\nu$$^{\dagger}$(mHz)	& $23.15 \pm 6.22$  & $8.6 \pm 2.6$   & $15.3 \pm 6.8$  & $12.0 \pm 4.5$  & $9.2 \pm 3.6$   & $51.8 \pm 31.7$  \\
\\
\hline
\\
$\chi^2$/dof            			& 137/150    	  &  381/338         &  310/269        & 442/434          & 317/294          & 53/62            \\
(continuum$^{b}$)			& (181/153)  	  & (409/341)  	     & (329/272)       & (478/437))       & (338/297)        & (78/65) \\
\\
Significance 					& $>$ 5$\sigma$  & $\approx$ 3.9$\sigma$ & $\approx$ 3$\sigma$  & $\approx$ 5$\sigma$ & $>$ 3$\sigma$ & $\approx$ 3.9$\sigma$ \\
\hspace{0.4cm}(ftest)			   &		  &	             &		   &  		     &		     &		\\
\\
    \hline\hline
    \end{tabular}
}
\\
  \end{flushleft}
\vspace{-0.25cm}
{\scriptsize 
\tablenotemark{a}{The effective exposure used for extracting the power density spectra.}\\
\tablenotemark{$\ast$}{We fit the continuum with a power-law model described as follows: \\
\begin{center}
\begin{math} Continuum = A + B\nu^{-\Gamma}\end{math}
\end{center}
where, $\Gamma$ is the power-law index of the continuum.
}\\
\tablenotemark{$\dagger$}{We model the QPOs with a Lorentzian. The functional form is as follows: 
\begin{center}
\begin{math} QPO = \frac {N_{QPO}} {1 + \left(\frac {2(\nu - \nu_{0})} {\Delta\nu}\right)^{2} } \end{math}
\end{center}
where, $\nu_{0}$ is the centroid frequency and $\Delta$$\nu$ is the FWHM of the QPO feature.\\
\tablenotemark{b}{The $\chi^2$/dof for the continuum are shown in braces.}\\
\tablenotemark{c}{Owing to only 8.8 ks of available good time interval, the significance of the QPO in the pn data was only 3$\sigma$. To confirm the presence of the QPO, we extracted a power density spectrum from combined MOS data.}\\
}
}
\end{table*}

We estimated the individual average count rates of source 5 and M82 X-1 as follows. First, we estimated the total counts from a given source by integrating its best-fit PSF until the core radius. We then divide this by the total exposure time to calculate an average count rate. The formula for the count rate is therefore:

\[
	count rate = \frac{1}{T}\times\left( \int_{0}^{r_{0}} \frac{N}{\left[1+ \left(\frac{r}{r_0} \right)^2\right]^\alpha} 2\pi|r| \,dr\right)
\]

where r is the radial distance from the centroid of the source and is defined as:
\[ 
r = \sqrt{(x-x_0)^2 + (y-y_0)^2} 
\]
where ($x_{0}$, $y_{0}$) is the best-fit centroid position of a given source. N is the best-fit value of the normalization of a given source. T is the effective exposure time. The count rates of source 5 and M82 X-1 estimated with the method described above are shown in the second and the third columns of Table 1, respectively. In observation 0112290201, source 5 clearly dominates the overall X-ray flux from M82. However in the rest of the observations M82 X-1's flux is greater than the flux from source 5. To minimize the contamination, we only considered observations in which M82 X-1's flux is $\ga$ source 5 flux. This filtering criterion resulted in a total of five observations (excluding observation 0112290201) to test for the timing-spectral correlation. We present the timing (PDS analysis) and the spectral analysis (energy-dependent surface brightness modeling) of these datasets in the following section.

\section{Results}

\subsection{Timing analysis}
The following analysis was carried out primarily using the EPIC-pn data with events in the energy range of 3.0-10.0 keV. We used the standard Science Analysis System (SAS) version 12.0.1 to extract the filtered event lists and the light curves. The standard filters of {\it (FLAG==0)} and {\it (PATTERN$<$=4)} were applied to all the datasets. The source events were extracted from a circular region of 33'' centered around the brightest pixel in each observation. This particular radius value was chosen to include roughly 90\% of the light from the source (as estimated from the fractional encircled energy of the EPIC-pn instrument). The background events were extracted from a nearby circular region of radius 50'' and free of other sources. We also removed episodes of high background flaring from our analysis. 

We constructed PDS from each of the five observations. These datasets, excluding observation 0206080101, have not been analyzed earlier and became public only recently (December 7$^{th}$ 2012). The data from observation 0206080101 has already been analyzed by Mucciarelli et al. (2006) \& Dewangan et al. (2006). We reanalyzed this observation to provide a consistent study of all the available data. All the PDS are shown in Figure 2 and Figure 3. All the power spectra shown here are so-called Leahy normalized where the Poisson noise level is equal to 2 (Leahy et al. 1983). It is clear that the overall behavior of all the PDS is the same. The power rises below $\approx$ 70-400 mHz with evidence for a QPO in the range of $\approx$ 30-220 mHz; And essentially Poisson noise at higher frequencies. To quantify this behavior, we fit a power law to the continuum and a Lorentzian to model the QPO (Belloni et al. 2002). The mathematical representation of the model can be found within the index of Table 2. This model fits adequately in all the cases with reduced $\chi^{2}$ in the range of 0.9-1.2. The best-fitting model parameters (derived from a fit in the frequency range of 0.001 Hz - 2.0 Hz) for each of the observation are shown in Table 2. We also indicate the $\chi^2$/dof (degrees of freedom) values for each of the fits along with the $\chi^2$/dof corresponding to the continuum model (in braces). The change in the $\chi^2$ serves as an indicator of the statistical significance of the QPOs. 

The longest available EPIC-pn good time interval during the observation 0657801901 was only 8.8 ks. The significance (ftest) of the QPO detected in the PDS extracted from this short exposure was $\approx$ 3$\sigma$. Fortunately, long uninterrupted data of duration $\approx$ 24 ks each was available from the MOS detectors. Therefore, to confirm the presence of the QPO, we extracted a PDS from the combined MOS data. The QPO is clearly evident in the MOS data with a detection significance of $\approx$ 5$\sigma$. The 3-10 keV EPIC-pn and combined EPIC-MOS PDS are shown in the left and the right panels of Figure 3, respectively. Finally, we analyzed the PDS of the backgrounds from each of the six datasets (five pn and one MOS) and note that they all are consistent with a constant Poisson noise.

\subsubsection{Origin of the mHz QPOs}
As mentioned earlier, the source region of M82 X-1 -- used for constructing the PDS -- is contaminated by the nearby point sources. The major source of contamination is source 5 which can reach flux levels comparable to M82 X-1. Therefore it is a concern as to which source (M82 X-1 or source 5) produces the QPOs. Work by Feng \& Kaaret (2007) has clearly shown that the few$\times$10 mHz QPOs originate from M82 X-1. More specifically, they demonstrate that the 54 mHz QPO during the observation 0112290201 and the $\approx$ 120 mHz QPO during the observation 0206080101 originate from M82 X-1. Furthermore, Feng et al. (2010) used the high angular resolution observations by {\it Chandra} to construct a clean PDS of source 5. They find that in the frequency range of $\approx$ 30-220 mHz the PDS of source 5 is essentially noise (see Figure 1 of Feng et al. 2010), suggesting that the power spectral contamination by source 5 is negligible. It is therefore likely that all the QPOs reported here (36-210 mHz) originate from M82 X-1. 

To confirm that M82 X-1 is indeed the origin of the mHz QPOs reported here, we carried out the same analysis as Feng \& Kaaret (2007). For each observation, we divided the source region into two semi-circles, one containing the majority of the flux from M82 X-1 (region A of the top panel of Figure 4) and the other dominated by the flux from source 5 (region B of the top panel of Figure 4). We then extracted the PDS from each of these individual half-circles. The PDS using only events from region A and from region B of observation 0657802301 (with the 210 mHz QPO) are shown in the middle and the bottom panel of Figure 4, respectively. It is clear that the QPO is evident in region A which is dominated by flux from M82 X-1. We found this to be the case in all the five observations. This analysis suggests that M82 X-1 is indeed the source of the mHz QPOs.

\subsection{Spectral analysis: Energy-dependent surface brightness modeling}
Due to contamination by point sources within the PSF of EPIC data a clean energy spectrum of M82 X-1 cannot be extracted. Energy spectral modeling of the previous high resolution {\it Chandra} observations of M82 X-1 suggests that its X-ray spectrum can be modeled by a simple power-law (see Kaaret et al. 2006). Furthermore, work by Feng \& Kaaret (2007) indicates that the absorbing column towards the source does not change significantly between observations that are randomly spread in time. Therefore, assuming the 3-10 keV X-ray spectrum of M82 X-1 can be modeled with a simple power-law, its hardness ratio (say ratio of the count rates in 3-5 keV and the 5-10 keV bands) will suffice as an indicator of the energy spectral power-law index. Therefore, we extracted the hardness ratio from each of the five observations by first carrying out the surface brightness modeling -- using the procedure described in Section 2 -- in the soft (3-5 keV) X-ray band and then in the hard band (5-10 keV). The resolved soft and hard count rates of M82 X-1 are indicated in the second and the fourth columns of Table 3. The corresponding hardness ratios are also shown.


\begin{figure}[ht]

\begin{center}
\includegraphics[width=3.in, height=8.02in, angle=0]{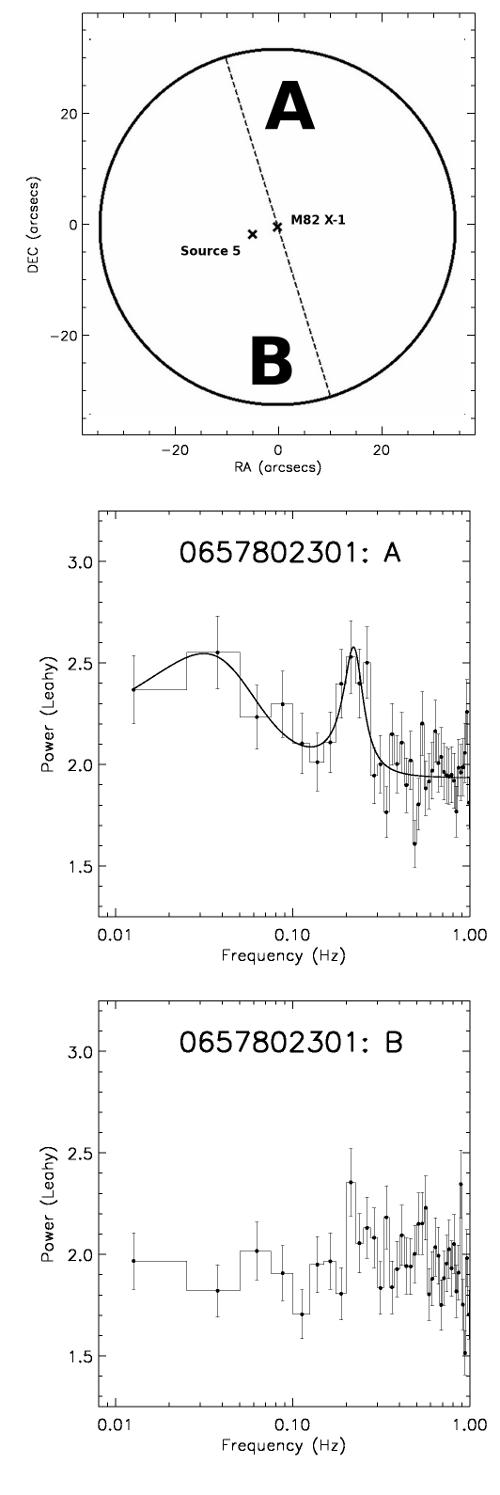}
\end{center}
\vspace{-.35cm}
{\textbf{Figure 4:} Top Panel: A circular source extraction region (radius of 33'' and centered on M82 X-1) demarcated as region A (not containing source 5) and region B (containing source 5). Similar to Feng \& Kaaret (2007), the dashed line is perpendicular to the line connecting M82 X-1 and source 5. Middle Panel: 3-10 keV EPIC-pn power density spectrum of region A. A best-fit model (bending power-law for the continuum and a Lorentzian for the QPO) is also shown ({\it solid}). Bottom Panel: 3-10 EPIC-pn power density spectrum of region B ({\it histogram}). This analysis shows that M82 X-1 is the source of the mHz QPOs.}
\label{fig:figure4}
\end{figure}


Furthermore, we ran simulations in {\it XSPEC} (Arnaud 1996) to constrain M82 X-1's spectral shape, i.e., the value of its power-law index. Our procedure is described as follows. First, using the {\it fakeit} command in {\it XSPEC}, we simulated a number of energy spectra (1000 in our case) each of which was described by a simple power-law modified by absorption, i.e., {\it phabs*pow} in {\it XSPEC}. We used the MOS1 responses generated using the {\it arfgen} and the {\it rmfgen} tasks for this purpose. These energy spectra spanned a wide range of power-law indicies (1-4) and normalizations (0.0001-0.01) with exposure time equal to the observing time of a given dataset. In essence, we generated a set of energy spectra as observed by EPIC-MOS1 and each prescribed by a power-law model with index and the normalization values in the range of 1-4 and 0.0001-0.01, respectively. From each of the five {\it XMM-Newton} observations (see Table 3), we calculated the MOS1 count rate of M82 X-1 in seven energy bands (3-10 keV, 3-9 keV, 3-8 keV, 3-7 keV, 3-6 keV, 3-5 keV and 3-4 keV) using surface brightness modeling technique described earlier. Within the suite of simulated spectra, we searched for the energy spectra whose count rates in the above bands are equal to the measured values (within the error bars) from surface brightness modeling of the real image. We find that the power-law index of M82 X-1 measured this way is only weakly constrained with a value in the range of 1.3-1.8. Note that this is consistent with the previous {\it Chandra} measurement of 1.67 (Kaaret et al. 2006).

\subsection{Timing-Spectral correlations}
The primary goal of the present work is to understand the nature of the mHz QPOs from ULX M82 X-1 by testing for a timing-spectral correlation similar to that seen in StMBHs with type-C LFQPOs. The basic  correlation that is characteristic of type-C LFQPOs in StMBHs is the dependence of the power-law index of the energy spectrum on the centroid frequency of the strongest QPO. Using all of the archival {\it XMM-Newton} observations we detected QPOs at five distinct frequencies from ULX M82 X-1 (see Section 3.1). Since a clean energy spectrum cannot be extracted with the present data we used the hardness ratio to represent the power-law index in each of these cases (see Section 3.2). Compiling all the results, we find that the hardness ratio shows no apparent dependence on the centroid frequency of the QPO. We find that as the centroid frequency of the QPO increases the hardness ratio appears to be constant. This is shown in the right panel of Figure 5. In addition, we plot the resolved MOS1 X-ray (3-10 keV) count rate of M82 X-1 against the centroid frequency of the QPO. We find a strong correlation with a Pearson's correlation coefficient of +0.97. We find that as the count rate of the source increases, the centroid frequency of the QPO also increases. This correlation is shown in the left panel of Figure 5. 


\begin{figure*}

\begin{center}
\includegraphics[width=6.5in, height=3.25in, angle=0]{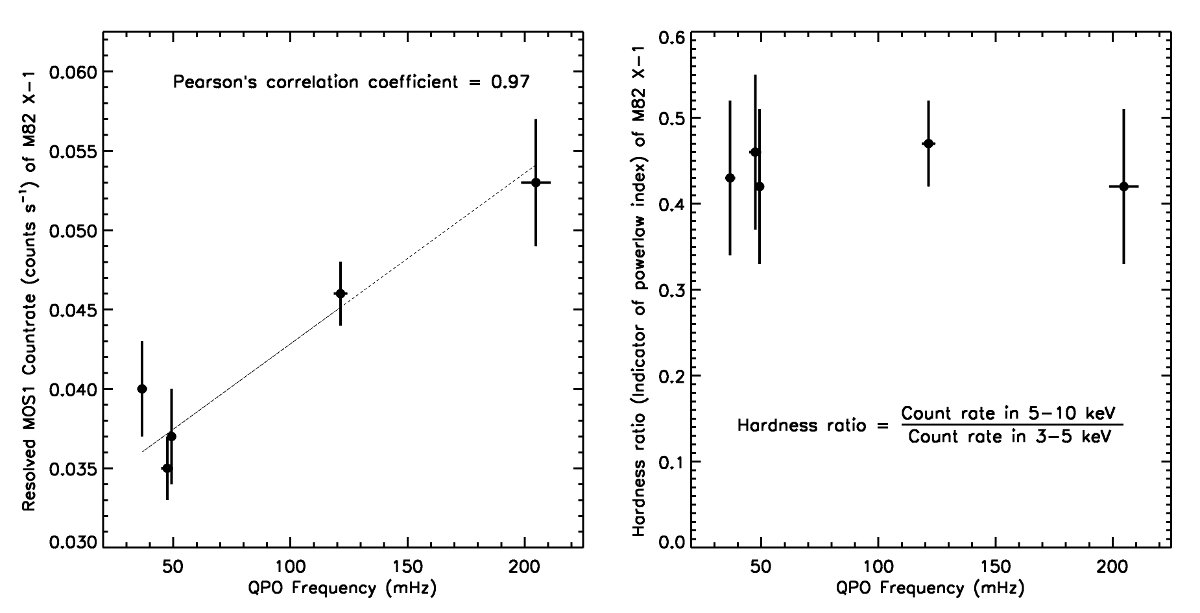}
\end{center}
{\textbf{Figure 5:} Timing-Spectral correlations. Left Panel: The correlation between the resolved MOS1 3-10 keV count rate of M82 X-1 ({\it Y-axis}) and the centroid frequency of the QPO ({\it X-axis}). The error bars are also shown. The value of the Pearson's correlation coefficient, which measures the significance of the correlation, is indicated at the top of the panel. The best-fit straight line ({\it dashed}) is also shown. Right Panel: The dependence of the hardness ratio of M82 X-1 ({\it Y-axis}) on the centroid frequency of the QPO ({\it X-axis}). The error bars are also shown. Using {\it XMM-Newton} data it is not possible to extract a clean energy spectrum, therefore, we use the hardness ratio instead which serves as an indicator of the power-law index of the energy spectrum. Compare with Figure 5 of Dheeraj \& Strohmayer (2012).}
\label{fig:figure5}
\end{figure*}


\begin{table*}
    \caption{ Summary of energy-dependent surface brightness modeling. We modeled all the MOS1 images in two energy bands: the soft (3-5 keV) and the hard (5-10 keV) X-ray bands. }\label{Table3}
{\small \center
    \begin{tabular}[t]{lcccccc}
    \hline\hline \\
ObsID\tablenotemark{a} & 3-5 keV\tablenotemark{b} & $\chi^{2}$/dof\tablenotemark{c} & 5-10 keV\tablenotemark{d} & $\chi^{2}$/dof\tablenotemark{e} & Hardness\tablenotemark{f} \\
	     		 & count rate (cts s$^{-1}$)  & 				  & count rate (cts s$^{-1}$)   & 		                     & ratio	 \\
	\\
    \hline \\
 0206080101 		 & $0.030 \pm 0.002$ 	      & 392/437 			  & $0.014 \pm 0.001$ 		& 233/437 			     & $0.47 \pm 0.05$ \\

	\\

 0657800101  		 & $0.026 \pm 0.003$ 	      & 352/437 			  & $0.011 \pm 0.002$ 		& 219/437 			     & $0.42 \pm 0.09$ \\ 

	\\

 0657801901  		 & $0.024 \pm 0.002$ 	      & 293/437 			  & $0.011 \pm 0.002$ 		& 140/437 			     & $0.46 \pm 0.09$ \\ 

	\\

 0657802101  		 & $0.028 \pm 0.003$ 	      & 362/437 			  & $0.012 \pm 0.002$ 		& 156/437 			     & $0.43 \pm 0.09$ \\ 

	\\

 0657802301 		 & $0.036 \pm 0.003$ 	      & 432/437 			  & $0.015 \pm 0.003$ 		& 280/437 			     & $0.42 \pm 0.09$ \\ 
\\
    \hline\hline
    \end{tabular}\\

\begin{flushleft}{
\tablenotemark{a}{The {\it XMM-Newton} assigned observation ID.}\\
\tablenotemark{b}{Resolved 3-5 keV MOS1 count rate of ULX M82 X-1. All the count rates are calculated using the formula described in the text (see Section 2).}\\
\tablenotemark{c}{The best-fit $\chi^2$/degrees of freedom (dof) from the surface brightness modeling using only the photons in the energy band of 3-5 keV.}\\
\tablenotemark{d}{Resolved 5-10 keV MOS1 count rate of ULX M82 X-1.}\\
\tablenotemark{e}{The best-fit $\chi^2$/degrees of freedom (dof) from the surface brightness modeling using only the photons in the energy band of 5-10 keV.}\\
\tablenotemark{f}{Hardness ratio of ULX M82 X-1 defined as the count rate in 5-10 keV over the count rate in 3-5 keV band.}\\}
\end{flushleft}
}
\end{table*}


\section{Discussion}
The so-called type-C LFQPOs of StMBHs are known to occur in the frequency range of $\sim$ 0.2-15 Hz. They are characterized by high quality factors (Q = centroid frequency/FWHM) of $\sim$ 7-12 and high fractional RMS amplitudes of $\sim$ 7-20\% (see Table 1 of Casella et al. 2005, Table 2 of Remillard et al. 2002 and Table 1 of McClintock et al. 2009). Another distinct feature of the type-C LFQPOs of StMBHs is that their centroid frequency is tightly correlated with the power-law index of the X-ray energy spectrum (Sobczak et al. 2000a; Vignarca et al. 2003). The relationship can be described as an increase in the power-law index with the QPO frequency with evidence for either a turn-over or constancy (saturation) beyond some higher value of the QPO frequency, i.e., beyond a certain high QPO frequency ($\sim$ 5-10 Hz) the power-law spectral index either decreases or remains constant (saturates) with increasing QPO frequency. The turn-over/saturation is known to hold over a small range ($\sim$ 5-15 Hz) of QPO frequencies (See Figure 10 of Vignarca et al. 2003). This general behavior has now been observed from various StMBHs including XTE J1550-564 (Sobczak et al. 2000a; Vignarca et al. 2003; Shaposhnikov \& Titarchuk 2009; McClintock et al. 2009),  GX 339-4 (Revnivtsev et al. 2001; Shaposhnikov \& Titarchuk 2009; Stiele et al. 2013), GRO J1655-40 (Sobczak et al. 2000a; Vignarca et al. 2003; Shaposhnikov \& Titarchuk 2009), Cygnus X-1 (Shaposhnikov \& Titarchuk 2007, 2009), H1743-322 (Shaposhnikov \& Titarchuk 2009; McClintock et al. 2009; Stiele et al. 2013), 4U 1543-475 (Shaposhnikov \& Titarchuk 2009) and GRS 1915+105 (Vignarca et al. 2003; Titarchuk \& Seifina 2009). While the slope of the correlation is different for different sources and sometimes different for the same source in a different outburst, the overall trend is the same. 

It is interesting to note that the hardness ratio of M82 X-1, an estimator of the energy spectral power-law index, remains constant over a wide range of QPO frequencies (36-210 mHz). There are two ways to interpret this result: (1) the mHz QPOs of M82 X-1 are indeed the analogs of type-C LFQPOs of StMBHs with the observed relationship representing the saturation portion of the trend or (2) the mHz QPOs of M82 X-1 are fundamentally different from the type-C LFQPOs of StMBHs as they show no apparent dependence on the power-law spectral index which is different from the positive correlation seen in StMBHs. Assuming the former to be the case, one can estimate the mass of the black hole in M82 X-1 by simply scaling the turn-over frequency of M82 X-1 ($\approx$ 40 mHz) to the turn-over frequency observed in various StMBHs ($\approx$ 5-10 Hz). Under the assumption that the turn-over frequency scales inversely with the mass of the black hole, the mass of the black hole in M82 X-1 can be estimated to be in the range of $\approx$ 500-1000 M$_{\odot}$, i.e., an IMBH. But on the other hand, saturation of the power-law index with the QPO frequency has never been seen over such a wide range of QPO frequencies in StMBHs. In StMBHs such a saturation is known to hold for QPO frequency changes of a factor of $\approx$ 1.5-3 (see Figure 10 of Vignarca et al. 2003; Shaposhnikov \& Titarchuk 2009). The QPOs observed from M82 X-1 occur in the frequency range of 36-210 mHz. This represents a factor of $\approx$ 6 change in the centroid frequency of the QPOs. Given such a large range in the QPO frequencies, it seems unlikley that the observed relationship represents the saturated portion of the type-C LFQPOs of StMBHs. In other words, \uline {the mHz QPOs of M82 X-1 may be fundamentally different compared to the type-C LFQPOs of StMBHs.} This is not surprising as similar dependence has now been seen from another ULX NGC 5408 X-1 (Dheeraj \& Strohmayer 2012).

Furthermore, mHz QPOs in the range of $\approx$ 2-300 mHz (a frequency range comparable to the QPOs of M82 X-1) have been observed from various StMBHs. These include GRO J0422+32 (QPOs with centroid frequencies of 300 mHz, 230 mHz and 200 mHz using Granat/SIGMA (40-150 keV), OSSE (35-60 keV) and BATSE (20-100 keV), respectively: Vikhlinin et al. 1995; Grove et al. 1998; van der Hooft et al. 1999), GRO J1719-24 (QPOs with centroid frequencies as low as 40 mHz and 300 mHz using BATSE (20-100 keV): van der Hooft et al. 1996), XTE J1118+480 (70-150 mHz QPOs detected using the USA experiment and RXTE: Wood et al. 2000; Revnivtsev et al. 2000), GX 339-4 (90-660 mHz QPOs using RXTE/PCA: Revnivtsev et al. 2001), GRO J1655-40 (100 mHz QPO using RXTE/PCA: Remillard et al. 1999), XTE J1550-564 (80-300 mHz QPOs using RXTE/PCA: Remillard et al. 2002; Cui et al. 1999), GRS 1915+105 (2-160 mHz QPOs using RXTE/PCA: Morgan et al. 1997), Cygnus X-1 (40-70 mHz QPOs using Granat/SIGMA: Vikhlinin et al. 1994) and H1743-322 (11 mHz QPO using RXTE and {\it Chandra}: Altamirano \& Strohmayer 2012). Moreover, the overall PDS of M82 X-1 show similarities with the PDS of GRS 1915+105 when it exhibits a few$\times$10 mHz QPOs and XTE J1550-564 when it shows a few$\times$10 mHz QPOs (compare Figure 2 \& 3 in this article with Figure 2 of Morgan et al. 1997 and Figure 2 of Cui et al. 1999). The continuum of the PDS of these three sources appear to be a simple power-law or a bending power-law. It is therefore possible that the mHz QPOs of M82 X-1 may be similar to the mHz QPOs of StMBHs and we are not able to observe the ``higher-frequency'' QPOs ($\sim$ 1-15 Hz) owing to very low count rate of M82 X-1 (Heil et al. 2009). If that were the case, the accreting black hole within M82 X-1 can be of stellar-mass. The large X-ray output may then be produced via some sort of a super-Eddington mechanism (see, for example, Begelman 2002). 

On the other hand it is interesting to note that the X-ray intensity of the source correlates with the QPO centroid frequency. Such a dependence has been observed from some StMBHs exhibiting type-C LFQPOs. These sources include XTE J1550-564 (see Figure 7 of Vignarca et al. 2003 and Table 1 of Sobczak et al. 2000b) and GRS 1915+105 (Figure 1 of Muno et al. 1999; Figure 1 of Reig et al. 2000; see Figure 2 \& 3 of Rodriguez et al. 2002). In addition, the constancy of the hardness ratio indicates that the energy spectral power-law index remains the same across these observations. Assuming that the 3-10 keV X-ray spectrum can be described by a simple power-law (previous high-resolution {\it Chandra} observations suggest this may be the case: see Kaaret et al. 2006) the X-ray count rate is directly proportional to the total X-ray/power-law flux. In which case, the left panel of Figure 5 is indicating a positive correlation between the X-ray/power-law flux and the QPO centroid frequency. 

Finally, we would like to point out that the implied spectral indicies are in the range of 1.3-1.8, which is within the range that LFQPO frequency increases with the spectral index in StMBHs (see, for example, Shaposhnikov \& Titarchuk 2009). In other words, there could be an increase in the spectral index of M82 X-1 with the mHz QPO frequency but the hardness ratio constraints are just not precise enough to show it. An effective way to know for certain if the QPO centroid frequency of ULX M82 X-1 is correlated or not correlated with its power-law spectral index is through joint {\it Chandra/XMM-Newton} observations; Where the {\it Chandra} data can be used to extract clean energy spectra of M82 X-1 and the {\it XMM-Newton} data can be used to estimate the QPO parameters of the source.

We would like to thank the anonymous referee for a careful review that has helped us improve this paper, and for pointing out that energy-dependent surface brightness modeling can be used to tackle the issue of contamination by the nearby source. We would also like to thank Dr. Margaret Trippe and Dr. Richard Mushtozky for their valuable comments. 

\vfill\eject

\end{document}